\documentclass[twoside,a4paper,11pt]{sea10}
\usepackage{graphicx}
\usepackage{hyperref}
\usepackage{movie15}
\begin{document}
\pagenumbering{arabic}
\pagestyle{myheadings}
\thispagestyle{empty}
{\flushleft\includegraphics[width=\textwidth,bb=58 650 590 680]{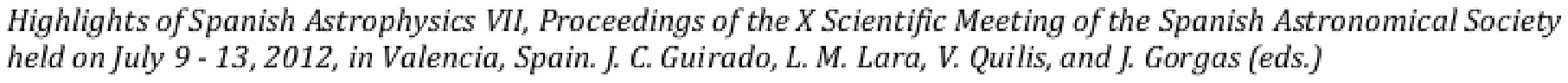}}
\vspace*{0.2cm}
\begin{flushleft}
{\bf {\LARGE
%
Solar sources of the geoeffective events in September 2011. 
%
}\\
\vspace*{1cm}
%
Judith Palacios$^{1}$,
Antonio Guerrero$^{1}$, 
Consuelo Cid$^{1}$, 
Elena Saiz$^{1}$, 
and 
Yolanda Cerrato$^{1}$
%
}\\
\vspace*{0.5cm}
%
$^{1}$
SRG-Spaceweather, Universidad de Alcal\'a, Spain 
%
\end{flushleft}
%
\markboth{
Solar sources of geoeffective events
}{ 
%
Palacios et al.
%
}
\thispagestyle{empty}
\vspace*{0.4cm}
\begin{minipage}[l]{0.09\textwidth}
\ 
\end{minipage}
\begin{minipage}[r]{0.9\textwidth}
\vspace{1cm}
\section*{Abstract}{\small
%
We investigate the geoeffective events happened from 8 to 20 Sept 2011, analysing the most plausible solar sources of these events, where coronal mass ejections and coronal holes play a fundamental role. The physical properties of the coronal holes, such as area and magnetic field, are studied through the Solar Dynamics Observatory instruments: AIA 193 \AA~images and HMI longitudinal magnetograms. The active regions which are the origin of the coronal mass ejections are analysed on AIA and SoHO-LASCO data.
%
\normalsize}
\end{minipage}
%
%
%
\section{Introduction \label{intro}}
Geomagnetic storms are characterized by a disturbance of the geomagnetic field. The most common way of identifying a geomagnetic storm is measuring the horizontal component of the magnetic field at low latitudes (through the geomagnetic indices $Dst$ or SYM-H) and establish a threshold in which these perturbations are considered as intense geomagnetic storms (if $Dst$ is lower than --100 nT; moderate when $Dst$ is lower than --50 nT, \cite{gonzalez94}). One of the effects of the storms on Earth is the intensification on northern lights occurrence, among others. 

The solar causes of the geomagnetic storms can be several: from the extremely dynamic coronal mass ejections (hereafter CMEs), the eruptive events  which ejects material off the corona; to the coronal holes (hereafter CHs), darker regions of open magnetic field lines where matter can escape, producing the fast solar wind. The ICMEs are the continuation of a CME across the interplanetary medium, and the high-speed streams (HSS) originate from CHs. In this work, we study the Sun-Earth chain backwards, from interplanetary data to find the solar drivers of the geoeffective events in September 2011.

\section{Data}
The data for the study of the CMEs are obtained from the LASCO catalog. The coronal images are obtained by SDO-AIA, while the line-of-sight magnetograms are provided by SDO-HMI.
The interplanetary data are retrieved from the missions ACE and WIND and SYM-H through the OMNIWeb site. The geomagnetic indices datasets have been stored for analysis at the spaceweather monitor website of the Universidad de Alcal\'a. 
\href{http://spaceweather.uah.es}{http://spaceweather.uah.es}.

\subsection{Interplanetary features}
\begin{figure}
\center
\includegraphics[scale=0.55]{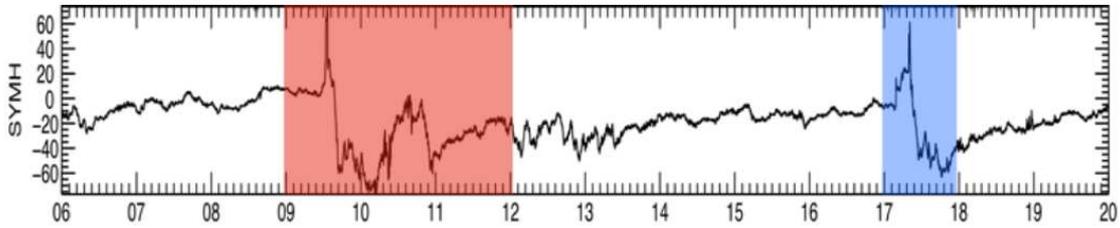} ~
\caption{ SYM-H data, showing the two geomagnetic storms on the time interval 6-20 Sept 2011. }\label{fig1}
\end{figure}

 We have identified two moderate geomagnetic storms in September 2011. Fig.~\ref{fig1} shows the geo\-magnetic index SYM-H (which is an indicator of how disturbed the magnetosphere is at low latitudes) with 1-min resolution. The plot shows the whole period (14 days) of SYM-H index, where both storms are indicated by peaks reaching lower than --60 nT. The analyses of the interplanetary data -- not shown here -- at Lagrangian point L1 (which is out of the magnetosphere and in the Sun-Earth line) are fundamental for characterising the features. A more comprehensive analysis of the interplanetary data of these storms is shown in the poster authored by Guerrero et. al, 2012 in this Proceeding volume; therefore we refer the reader for details when mentioning interplanetary features.

In Fig.~\ref{fig1}, the first storm is marked by the area in red from Sept 9 to 12, where SYM-H shows two consecutive peaks indicating a long duration storm not seen in common geomagnetic storms. The first peak (SYM-H peak  $\sim$ -- 60 nT) is associated to an ICME while the second one (SYM-H peak  $\sim$ -- 70 nT) is caused by the interaction region between the ICME, the background solar wind and the HSS which extends all the way to the second storm. The CH in this study is the source of this HSS and is clearly seen by high values of the flow speed $(FS)$ and fluctuating magnetic field ($B_{z}$) and plasma temperature $(T)$ filling the region between both storms in the interplanetary medium. The second storm, which is marked on blue, is caused by a compressed ICME. The response of the magnetosphere did not recovered (negative SYM-H) completely when the second ICME arrives to Earth due to negative and fluctuating $B_{z}$. 

\subsection{Associated coronal mass ejections} 

\begin{figure}
\center
\includegraphics[scale=0.40]{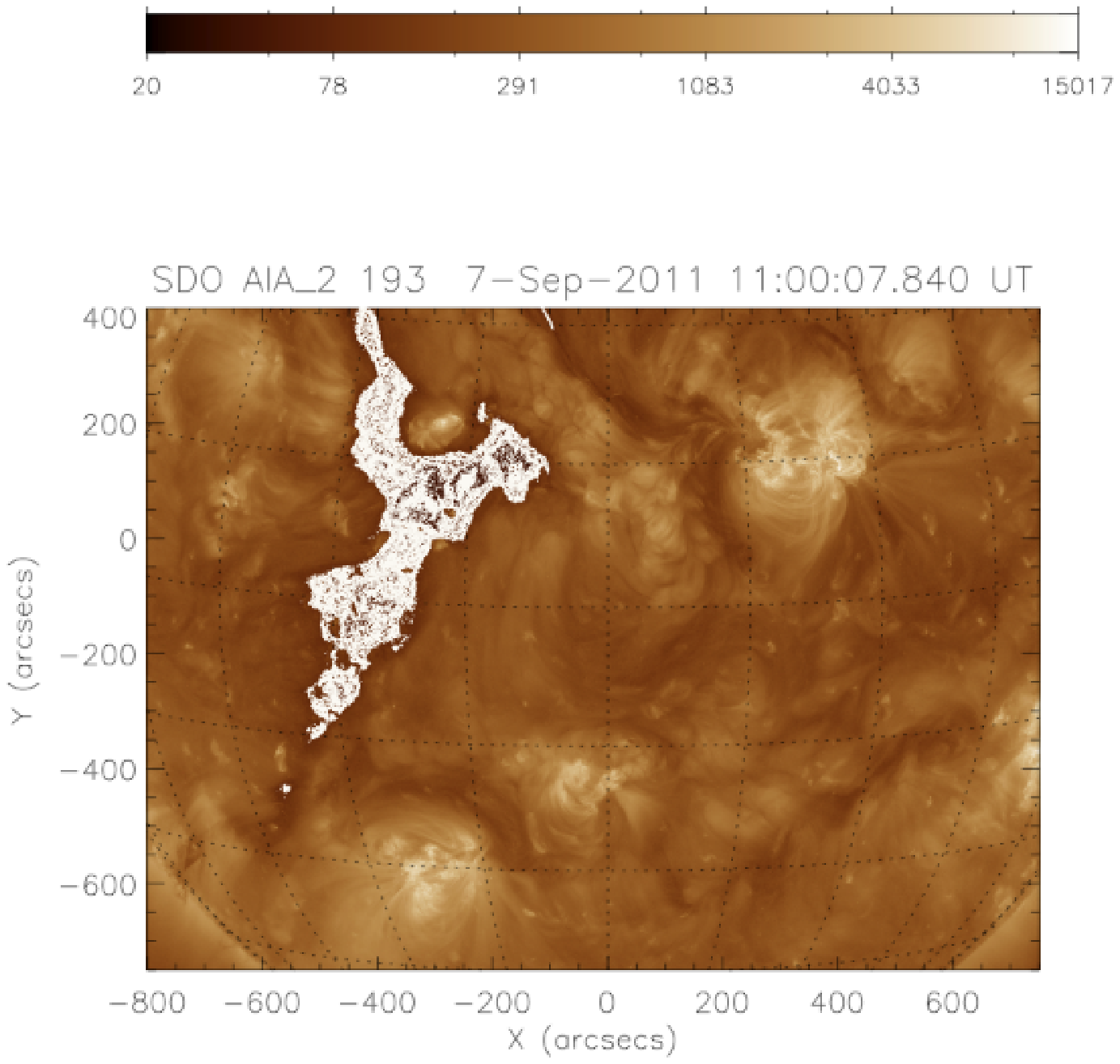} ~
\includegraphics[scale=0.40]{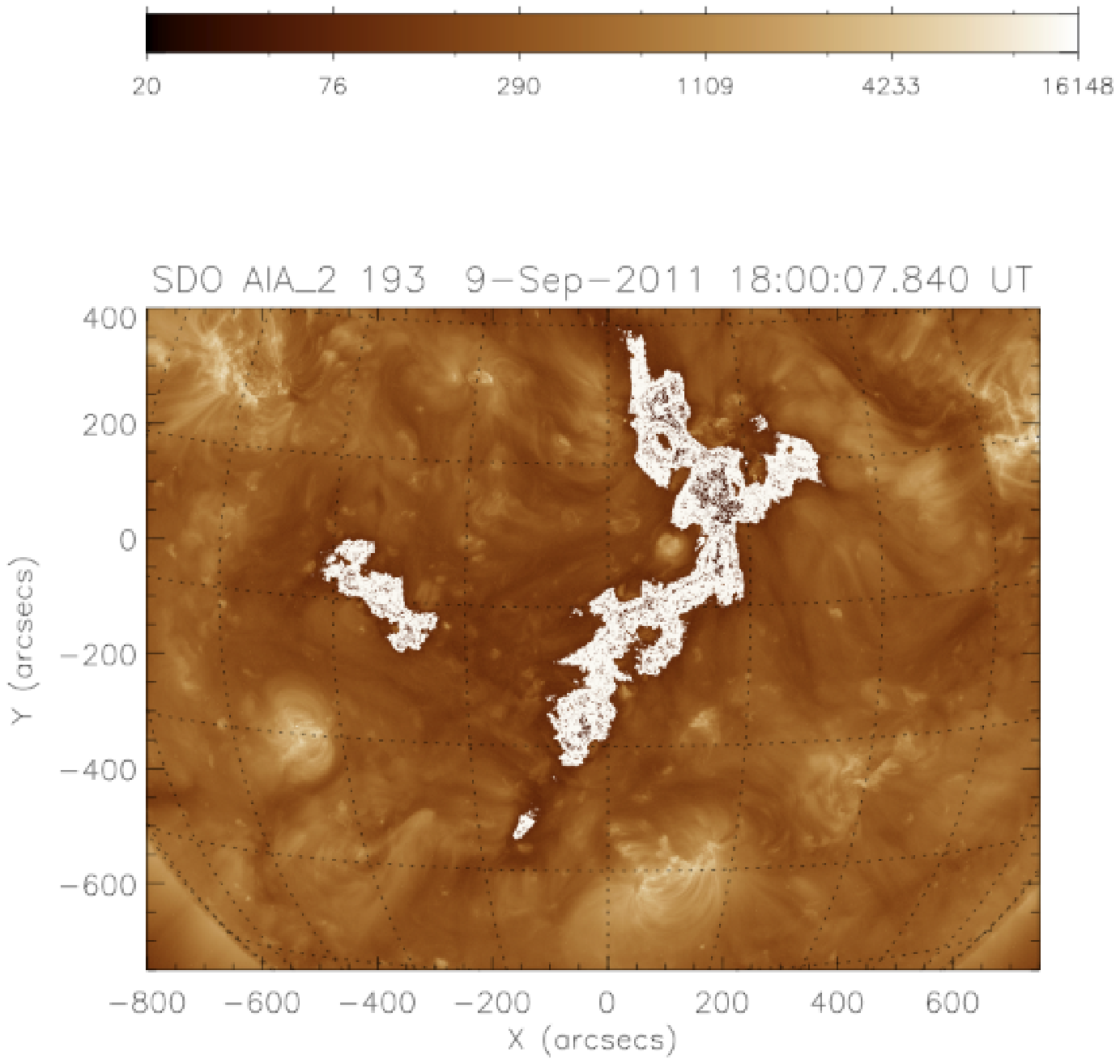} ~\\
\includegraphics[scale=0.40]{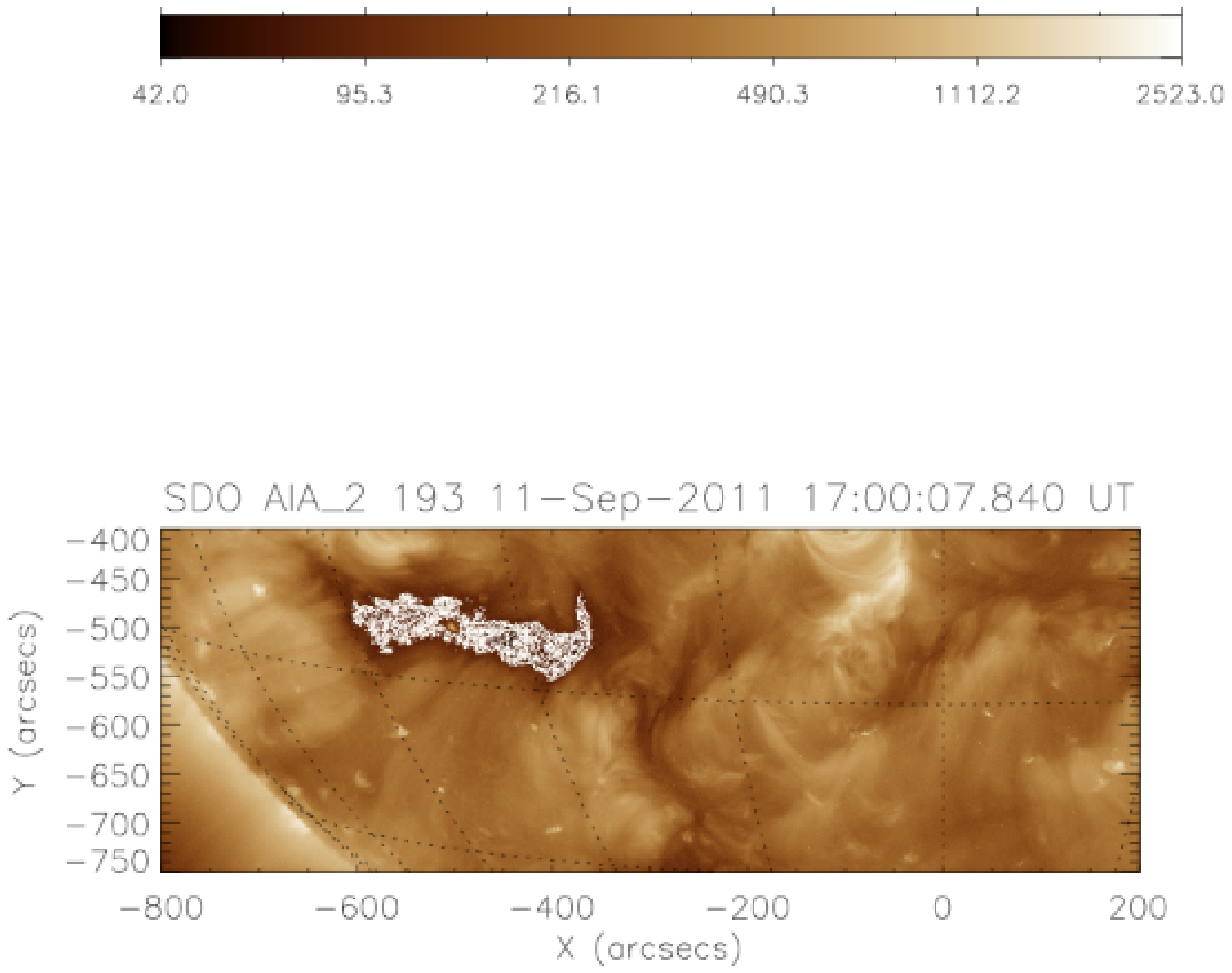} ~
\includegraphics[scale=0.40]{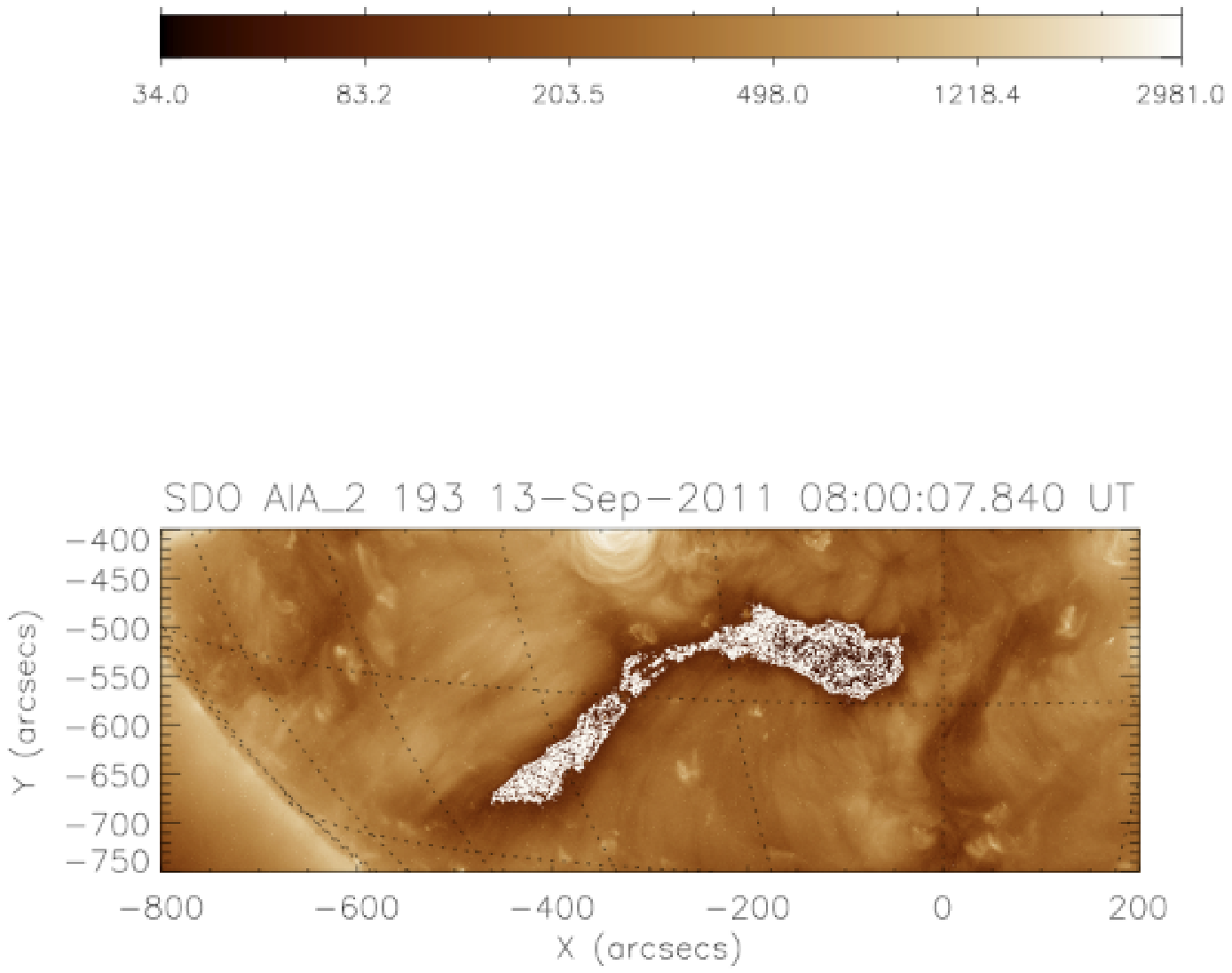} 
\caption{$\it{Top}$: CH1 feature, marked in white, before and after the appearance of the coronal dimming (located at the left of the main CH). $\it{Bottom}$: CH2 evolution, from the initial to the maximum area, before CME2. The colorbars on top mark the intensity in counts. }
\label{fig2} 
\end{figure}

The velocities, timing and characteristics of the CMEs in this Section are taken from the LASCO catalog (\href{http://cdaw.gsfc.nasa.gov/CME\_list/}{http://cdaw.gsfc.nasa.gov/CME$\_$list/}). The chosen candidates were selected by their plane-of-the-sky speed and the active region (AR) position from where they were ejected. The CME related to the first geomagnetic storm (hereafter CME1) was ejected on 2011, September 6  at 23:06 UT, reaching a plane-of-the-sky speed of 575 km~s$^{-1}$. It was a full halo CME\footnote{This type of CME is the most likely to generate a geomagnetic storm \cite{venka2003}} originated at the Active Region (AR) AR 11283, located on the solar West, coordinates N14W28. The related flare was categorised as X2.1 class.

\begin{figure}[ht]
\centering
 \includegraphics[scale=0.7]{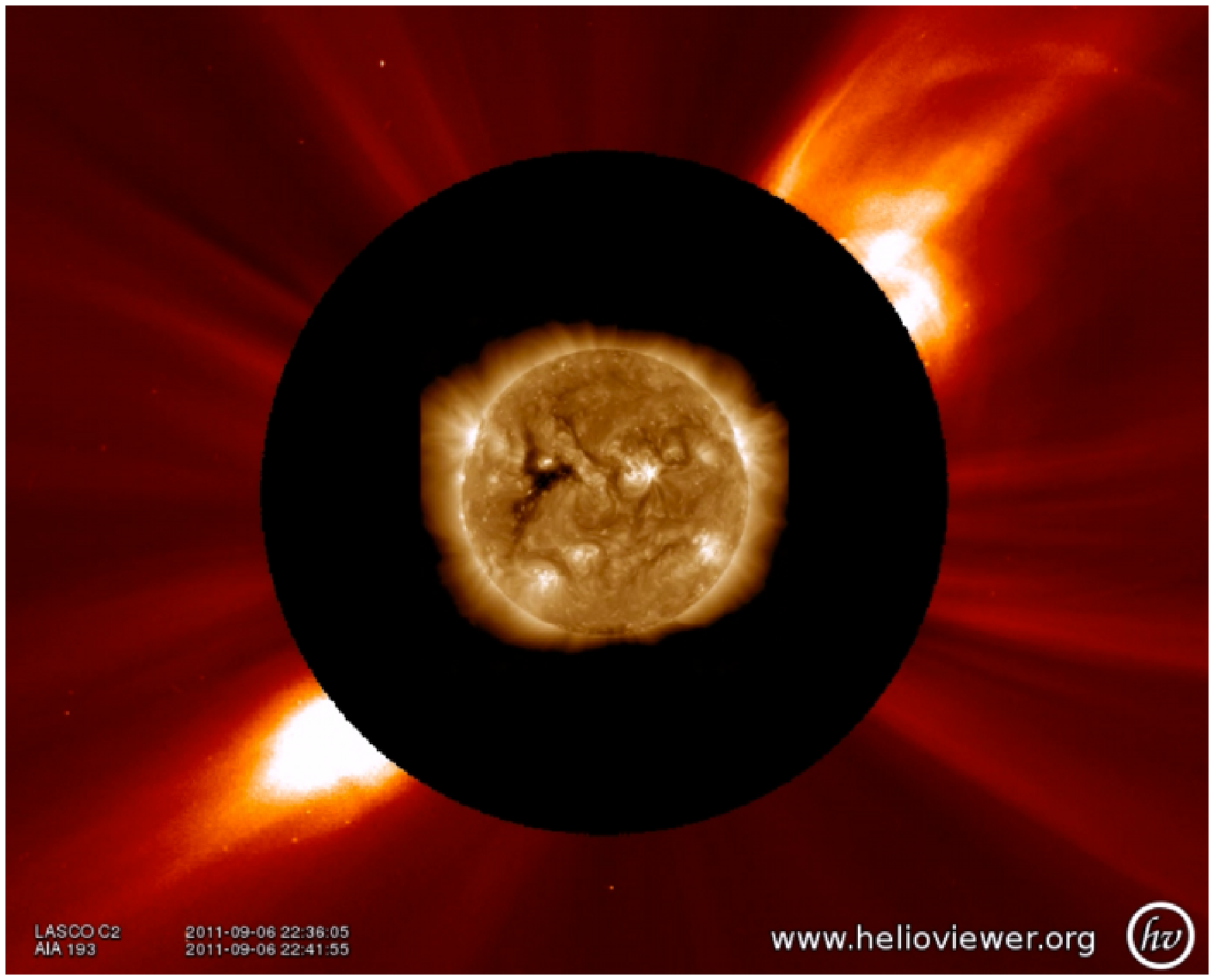}
\caption{Multimedia material: LASCO C2 and AIA 193 \AA~ movie. The frame is chosen to show the flare and the CME start in Sept 6.}\label{mov}
\end{figure}

The second CME (hereafter CME2) was released on September 14 at 00:05 UT from the AR11292 (solar East, coordinates N22E62) as a partial-halo CME, with a plane-of-the-sky speed of 408 km~s$^{-1}$, and related to a minor flare (C2.9). The CMEs and CH evolution are displayed at the multimedia material embedded in Fig.~\ref{mov}.

\subsection{Physical properties of the coronal holes} 

\begin{figure}
\center
\includegraphics[scale=0.35]{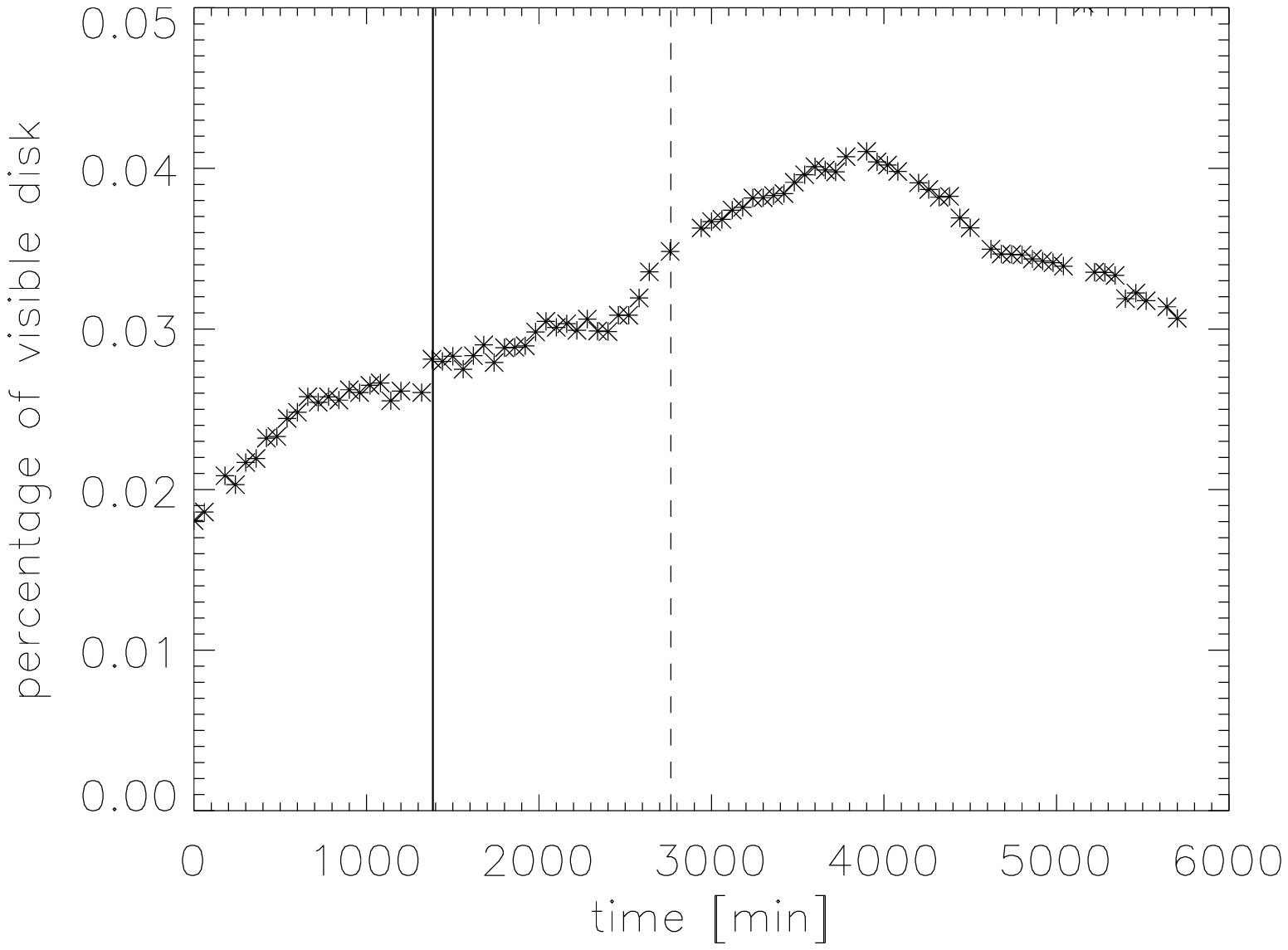} ~
\includegraphics[scale=0.35]{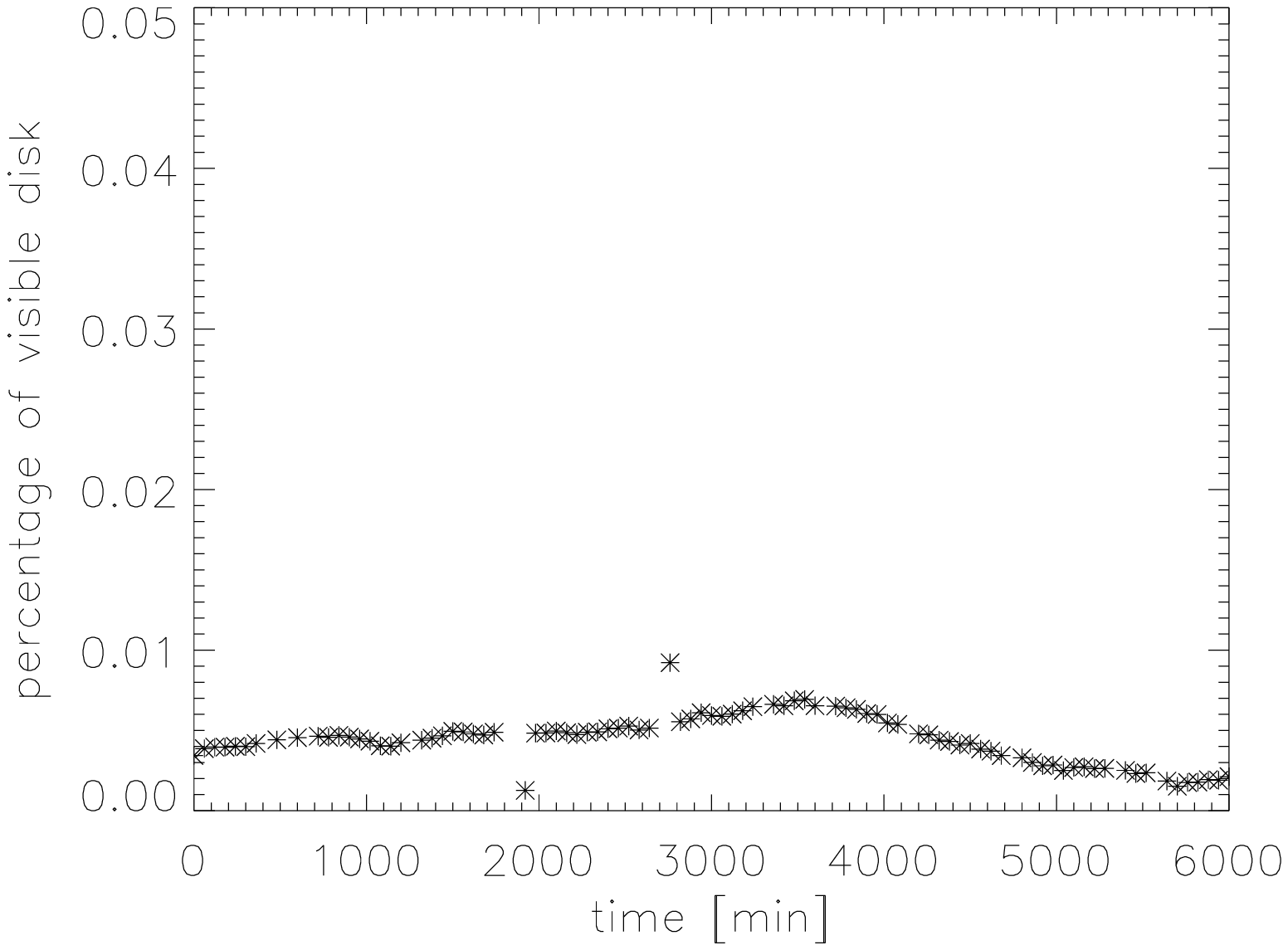}
\caption{CH1 and CH2 area. $\it{Left}$: The increase of area on CH1 after CME1 (marked by a solid line) and the CME provoking the dimming (dashed line) is around 30\%. The point $t=0$ on the plot corresponds to 2011 Sept 6, 00:00 UT. $\it{Right}$: CH2 area is measured from 11 (00:00 UT) to 14 Sept. }
\label{fig3}
\end{figure}

The coronal hole area is measured through simple thresholding on AIA 193 \AA~ images and corrected for foreshortening. This wavelength corresponds to 1.4 MK \cite{brosius2012}, although AIA 211 \AA~ is also suitable for this purpose. The HMI longitudinal magnetograms are used to calculate the mean magnetic field strength on the area covered by the coronal hole. 

The coronal hole related with the first storm (hereafter CH1) extends approximately from latitude --25$^\circ$ to 30$^\circ$ and 15$^\circ$ wide. This large CH increases its surface slightly due to the first CME. However, the most conspicuous area increase (a large coronal dimming shown in Fig.~\ref{fig2}, top right) is related to another CME produced on Sept 7 at 23:06 UT, with a plane-of-the-sky speed of 792~km~s$^{-1}$ and linked to a X1.8 flare. Unfortunately, due to the relatively low speed it is unlikely that this CME may be the source of the first geomagnetic storm. 

As shown in Fig.~\ref{fig3} (left), the area of the CH1 increases after the CME by the coronal dimming around 30\%. Another important physical property of the CHs, the magnetic field strength, can be checked not only on the Sun's surface, but in the interplanetary medium data. We verified that the polarities of the interplanetary magnetic field are consistent to the solar surface ones. The mean magnetic field strength measured on photospheric magnetograms is negative ($\sim$ --5G), and the dimming contributes with positive mean magnetic strength. The total photospheric magnetic flux of the corresponding area is $\sim$ 10$^{21}$ Mx, after removing the noisy pixels with a 3$\sigma$ mask. These value ranges are in agreement to \cite{meunier2005, zhang2006}.

The area and magnetic field strength of CH2 remains more constant (Fig.~\ref{fig3}, right), since no dimming affected the characteristics of this feature. The mean magnetic field strength is $\sim$ --2G, and the corresponding photospheric magnetic flux is similar to CH1.  

\section{Discussion and concluding remarks}

We have studied the CMEs and CHs that led to two moderate geomagnetic storms in Sept 2011. The combination of the CME and coronal hole can be highly geoeffective, since it can provoke coronal dimmings or other interactions, as CME deflections \cite{gopalswamy2009}. The boundaries of coronal holes play an important role as it is the location where fast wind high-speed streams originate by reassembling open and closed magnetic field lines, leading to reconnection \cite[e.g.]{mardjaska2004} and thus reconnection features appear, as i.e. X-ray jets\cite{morenoinsertis2008}.  The interplanetary medium feature ICMEs and interacting structures characterized by negative and fluctuating magnetic field which originated the geomagnetic storms. 
%
%
\small  
%
\section*{Acknowledgments}   
%
We would like to thank to NASA, Goddard Space Flight Center to make available ACE, WIND and SYM-H data via OMNIWeb and SDO (Solar Dynamics Observatory) data, as well as Naval Research Laboratory for the EIT and LASCO images from SoHO, and also STEREO Science Center. The LASCO CME catalog is generated and maintained at the CDAW Data Center by NASA and The Catholic University of America in cooperation with the Naval Research Laboratory.  We also want to thank the Community Coordinated Modeling Center (CCMC) for the simulation codes available, and the website and tool Helioviewer for composing the attached movie.

This work has been supported by grant AYA2009-08662 from the ``Ministerio de Ciencia e Innovaci\'on'' of Spain and grant PPII10-0183-7802 from ``Junta de Comunidades de Castilla- La Mancha''.


%


\begin{thebibliography}{}
\small
%

\bibitem{brosius2012}{ Brosius, J.~W. \& Holman, G.~D., 2012, A\&A, 540, 24 }
\bibitem{gonzalez94}{Gonzalez, W.~D., Joselyn, J.~A., Kamide, Y., Kroehl, H.~W., Rostoker G., Tsurutani, B.~T.,  and Vasyliunas, V.~M., 1994, J. Geophys. Res., 99(A4), 5771}
\bibitem{gopalswamy2009}{Gopalswamy, N., M{\"a}kel{\"a}, P.,  Xie, H.,  Akiyama, S., Yashiro, S., 2009, JGRA, 114, A00A22}
\bibitem{mardjaska2004}{Mardjaska, M.~S., Doyle, J.~G., Van Driel-Gesztelyi, L., 2004, ApJ., 603, L57}
\bibitem{meunier2005}{Meunier, N., 2005, A\&A, 443, 309}
\bibitem{morenoinsertis2008}{Moreno-Insertis, F.; Galsgaard, K.; Ugarte-Urra, I., 2008, ApJ., 673, 211}
\bibitem{venka2003}{Venkatakrishnan, P., \& Ravindra, B., 2003, Geophys. Res. Lett., 30, 2181}%
\bibitem{zhang2006}{Zhang, J., Ma, J., and Wang, H., 2006, ApJ.,  649, 464  }

%
%
\end{thebibliography}
\end{document}